# A modified J model for efficiently calculating the electromagnetic fields of ReBCO no-insulation pancake coils using an explicit-implicit hybrid algorithm

Yunkai Tang[1,2], Donghui Liu[1,2], Shouhong Shan[1,2], Dongke Li[1,2], Xiaohui Lin[3,4], Shuiliang Zhen[3,4], Chao Zhang[3,4], Huadong Yong*[1,2]

*1. Key Laboratory of Mechanics on Disaster and Environment in Western China, Ministry of Education of China, Lanzhou University, Lanzhou 730000, China*

*2. Institute of Superconductor Mechanics, Lanzhou University, Lanzhou 730000, China*

*3. Shanghai Superconductor Technology Co., Ltd, Shanghai 201203, China*

*4. Shanghai Key Laboratory of High-Temperature Superconductor Materials and Application, Shanghai 201203, China*

**Abstract**

Rare-earth $(Re)Ba_2Cu_3O_{7-x}$ (ReBCO) no-insulation (NI) coil is widely concerned due to its excellent electromagnetic and thermal properties. However, the presence of the turn-to-turn shunts in NI coils leads to that complexity of numerical simulation is increased. In this paper, a modified J model is proposed and the corresponding explicit-implicit hybrid algorithm is designed to calculate NI coils. The numerical results are in good agreement with the experimental data and the circuit model. The homogenization model is also proposed to simulate the large-scale NI coils in the background magnets. The modified J model has good accuracy and fast calculation speed, which can also be used to solve electromagnetic fields of insulation coils efficiently.

homogenization model.

Email: yonghd@lzu.edu.cn

# 1 Introduction

Second-generation (2G) high-temperature superconducting (HTS) rare-earth (Re) $Ba_2Cu_3O_{7-x}$ (ReBCO) coated conductor has been the strong candidate for the fabrication of the ultra-high-field magnets due to its excellent current-carrying capacity [1, 2]. A variety of numerical methods for calculating the electromagnetic fields of ReBCO magnets have been proposed, such as H, T-A, H-Φ, J-A formulations [3-10], and so on. All of them solve the partial differential equations (PDEs) with different unknown physical quantities by the finite element method (FEM). The air domain must be used to impose boundary conditions, which increases the amount of computation. However, the air domain can be removed in several methods, such as the approach with the variable of the current density **J**. The reason is that the relationship between current density **J** and the magnetic vector potential **A** satisfies Poisson's equation, which means that **A** can be represented by the integration of **J** and Green's function over the whole domain [11]. Since **J** is zero in the air domain, we only need to calculate the electromagnetic field in the superconductor domain. The approaches with J as the variable were proposed by Prigozhin and Brandt in 1996 [12, 13]. The former derived the functional of the superconductor, that is, **J** is the optimal solution of the functional. Then, Pardo *et al* extended this method as the minimum electromagnetic entropy production (MEMEP) method and successfully simulated the electromagnetic fields of various superconducting structures [14-19]. Especially, the method of functional optimization is convenient when the Bean critical state is considered since the functional is independent on the time, and the electromagnetic field of superconductors is described as a convex quadratic programming problem with equality and inequality constraints [20, 21]. However, the Lagrange multiplier vector is added to impose current constraints (i.e., the current continuity equation) [22] and may lead to the increase of the degrees of freedom. In Brandt's approach, the integral equation with time derivatives is solved directly. By spatial discretization, the integral equation is transformed into a system of ordinary differential equations (ODEs) with the variable **J**. Yazawa et al calculated the AC loss of HTS by Brandt's method [23]. However, another unknown quantity, $\nabla \Phi$, needs to be solved in the charging process. A simple treatment similar to the penalty method was used by expressing $\nabla \Phi$ as the current constraint with a parameter $\gamma$ to simulate HTS with a constant critical current density by Otten and Grilli [24], where $\gamma$ is set as 100 V/Am experientially. Later on, Lai et al eliminated $\nabla \Phi$ by expressing the current constraint as the time derivative form [25], which is easier to be implemented and Brandt's method is named as J model [26].

The above methods can effectively calculate the electromagnetic field of the insulation coils, whose current density only has the hoop component $J^\theta$. In 2011, Hahn et al proposed the no-insulation (NI) technology, and it means that the radial shunt $J^r$ can be achieved by low turn-to-turn contact resistance [27]. Thus, the thermal stability of coils is improved remarkably [28]. However, the calculation of electromagnetic fields of NI coils becomes a challenge problem. The usual method is to construct the equivalent circuit model based on the Kirchhoff's law [27, 29]. It can calculate the hoop current $I^\theta$ and the radial current $I^r$ in each turn, and describe the delay characteristic of NI coils accurately. However, in order to obtain the electromagnetic field distribution, $I^\theta$ needs to be imported into H or T-A formulation. The partial element equivalent circuit model [30], the equivalent circuit grid model [31] and the three-dimensional (3D) circuit model [32, 33] were also proposed to solve the problem. Recently, Zhou *et al* and Pardo *et al* respectively extended T-A formulation [34] and MEMEP [35], and the NI coils can be characterize with the modified methods. However, it is still difficult to deal with NI coils with the J model. Since J model has a higher computational efficiency, it is meaningful to extend the J model to solve the NI coils.

Based on the above discussions, a modified J model is proposed to simulate the electromagnetic fields of ReBCO NI coils in this paper. Compared to the previous method, the modified J model is easier to be implemented. We eliminate the unknown quantities $\nabla \Phi$ and $J^r$ by combining the current constraints with the hoop and radial voltage relations, so that the unknown quantity can be directly solved by $J^\theta$, which reduces the degrees of freedom as much as possible. As $J^\theta$ is obtained, $J^r$ can be solved by the system of linear current constrain equations. In order to settle the stiffness of the ODEs, an explicit-implicit hybrid algorithm is designed so that the calculation can be stably advanced with a large time step. Moreover, the homogenization model is also proposed to calculate the large-scale NI magnets. The results calculated by our method are in good agreement with those obtained by the experiment and the circuit model, and the computation speed is fast. Finally, this model is also extended to calculate the electromagnetic field distribution of insulation coils by increasing the contact resistivity.

**2 Theories and Algorithms**

*2.1 J model for NI pancake coils*

Based on Maxwell's equations, the following relation holds [25]:

$$\partial_t \mathbf{A}_{in} + \partial_t \mathbf{A}_{ex} + \mathbf{E} + \nabla \Phi = \mathbf{0}, \tag{1}$$

where **E** is the electric field and $\Phi$ is the electric scalar potential. $\mathbf{A}_{ex}$ and $\mathbf{A}_{in}$ are the magnetic vector potentials generated by the background magnet and the current density **J**, respectively. The material properties of the conductor are described by the constitutive equation

$$\mathbf{E} = \rho \mathbf{J}. \tag{2}$$

In addition, it is also necessary to supplement the current continuity equation

$$\nabla \cdot \mathbf{J} = 0, \tag{3}$$

which is equivalent to Ampere's law ignoring the electrical displacement **D**.

For the NI pancake coils, we assume that the electromagnetic field is axisymmetric distribution. Therefore, Eq. (1) along the hoop direction is

$$\partial_t A_{\text{in}}^\theta + \partial_t A_{\text{ex}}^\theta + E^\theta + (\nabla \Phi)^\theta = 0. \tag{4}$$

The constitutive equation Eq. (3) along the hoop direction is

$$E^\theta = E_c \left( \frac{|J^\theta|}{J_c} \right)^n \frac{J^\theta}{|J^\theta|}, \tag{5}$$

where $E_c = 10^{-4}$ V/m and $J_c$ is the critical current density. A complete circuit is composed of the hoop branch $l_h$ and the radial branch $l_r$. Thus,

$$\oint \nabla \Phi \cdot d\mathbf{l} = \int_{l_h} (\nabla \Phi)^\theta \, dl + \int_{l_r} (\nabla \Phi)^r \, dl = 0. \tag{6}$$

Ignoring the magnetic field and electromagnetic induction generated by the radial current density $J^r$, it can be obtained as

$$(\nabla \Phi)^r = -E^r = -\rho^r J^r, \tag{7}$$

where $\rho^r$ is the equivalent radial resistivity, which is defined as the contact resistivity $\rho_{ct}$ divided by the tape thickness $d$. Note that the contribution of each tape component resistivity to $\rho^r$ is neglected. Therefore, Eq. (6) becomes [34]

$$2\pi r \cdot (\nabla \Phi)^\theta + d \cdot \rho^r J^r = 0. \tag{8}$$

Finally, Eq. (3) is formulated as the following current constraints:

$$I_{\text{tr}} = I_i^r + I_i^\theta = \int_{S_i} J^\theta \, dS + \int_{D_i} J^r \, dD. \tag{9}$$

where $I_i^r$, $I_i^\theta$, $S_i$ and $D_i$ are respectively the radial current, the hoop current, the cross-sectional area and the lateral area of the $i$-th turn, and $I_{\text{tr}}$ is the transport current.

Each tape's width $w$ is divided into $N_w$ meshes, and since the tape is very thin, only one mesh is used along its thickness direction. Therefore, the total number of meshes is $N = N_w \times N_t$. For simplicity, the mesh is divided uniformly, and we assume that $J^\theta$ in each mesh to be a constant. It can be proven that $J^r$ is independent on the coordinate $z$ if $\rho_{\text{ct}}$ is constant along the coordinate $z$ and the magnetic vector potential generated by $J^r$ is ignored. Therefore, the equations are discretized into the following system of differential-algebraic equations (DAEs):

$$\begin{cases} [M]\{\dot{J}^\theta\} = [C_1]^{-1}[L]^T[C_2]\{J^r\} - \{E^\theta\} - \{\dot{A}_{\text{ex}}^\theta\} \\ [A_1]\{J^\theta\} + [A_2]\{J^r\} = \{b_{\text{eq}}\} \end{cases}. \tag{10}$$

The left of the first equation in Eq. (10) is the discrete form of $\partial_t A_{\text{in}}^\theta$ and at the $i$-th mesh [25]

$$\partial_t A_{\text{in},i}^\theta = \frac{\mu \cdot \Delta S_j \cdot \partial_t J_j}{\pi k} \sqrt{\frac{r_j}{r_i}} \left[ K(k) \cdot \left(1 - \frac{k^2}{2}\right) - F(k) \right], \tag{11}$$

where $\mu$ is the permeability, $r_i$ is the radial coordinate of the $i$-th mesh, and $r_j$, $\Delta S_j$ and $J_j$ are the radial coordinate, the area and the current density of the $j$-th mesh. Therefore, each element of $[M]$ is

$$M_{ij} = \frac{\mu \cdot \Delta S_j}{\pi k} \sqrt{\frac{r_j}{r_i}} \left[ K \cdot \left(1 - \frac{k^2}{2}\right) - F \right]. \tag{12}$$

The parameter $k$ in Eqs. (11) and (12) is

$$k = \sqrt{\frac{4r_i r_j}{\sqrt{(r_i + r_j)^2 + (z_i - z_j)^2}}}, \tag{13}$$

and $K$ and $F$ are respectively the first and second complete elliptic integrals. Other matrixes are

$$\{b_{eq}\}_{(N_t \times 1)} = \begin{Bmatrix} I_{tr} \\ I_{tr} \\ \vdots \\ I_{tr} \end{Bmatrix} \quad [A_1]_{(N_t \times N)} = \begin{bmatrix} \overbrace{\Delta S_1 \cdots \Delta S_{N_w}}^{N_w} & & \\ & \ddots & \\ & & \overbrace{\Delta S_{N-N_w+1} \cdots \Delta S_N}^{N_w} \end{bmatrix}$$

$$[A_2]_{(N_t \times N_t)} = \begin{bmatrix} 2\pi R_1 w & & & \\ & 2\pi R_2 w & & \\ & & \ddots & \\ & & & 2\pi R_{N_t} w \end{bmatrix} \quad [C_1]_{(N \times N)} = \begin{bmatrix} \overbrace{2\pi R_1}^{N_w} & & \\ & \ddots & \\ & & \overbrace{2\pi R_{N_t}}^{N_w} \\ & & \ddots \end{bmatrix}$$

$$[C_2]_{(N_t \times N_t)} = \begin{bmatrix} d\rho^r & & & \\ & d\rho^r & & \\ & & \ddots & \\ & & & d\rho^r \end{bmatrix} \quad [L]_{(N_t \times N)} = \begin{bmatrix} \overbrace{1 \cdots}^{N_w} & & \\ & \overbrace{1 \cdots}^{N_w} & \\ & & \ddots \\ & & & \overbrace{1 \cdots}^{N_w} \end{bmatrix}, \quad (14)$$

where $w$ is the width of the tape, and $R_i$ is the radius of the $i$-th turn. Then, Eq. (10) becomes a non-linear system of ordinary differential equations (ODEs)

$$[M]\{\dot{J}^\theta\} = [C_1]^{-1}[L]^T[C_2][A_2]^{-1}(\{b_{eq}\} - [A_1]\{J^\theta\}) - \{E^\theta\} - \{\dot{A}^\theta_{ex}\}, \quad (15)$$

where the non-linear term is $\{E^\theta\}$ from Eq. (5), and the initial condition is

$$\{J^\theta\}^{(0)} = \{J^\theta\}\big|_{t=0} = \{0\}. \quad (16)$$

*2.2 Explicit-implicit hybrid algorithm*

In the original J model, the forward difference formula (FDF) was used in an explicit format to calculate infinitely long ReBCO tapes and insulating coils [25]. To prevent the calculation from diverging caused by the E-J non-linear constitutive equations, the authors also proposed a kind of variable step size technique [25]. However, for no-insulation coils, the explicit algorithm requires a very small step size to ensure stability, which needs a long time to solve the electromagnetic response. The reason is that the change rates of each element of $\{J^\theta\}$ are different, so that Eq. (15) is stiffer than the original J model equation, which needs a smaller step to ensure the stability of the algorithm. The stability analysis of the algorithm is complex and can be found in textbooks of numerical analysis [36].

For the above reasons, the implicit algorithm becomes the necessary choice, and we use the implicit solvers built into the commercial software MATLAB. To improve the computational efficiency, it is necessary to provide the mass matrix $[M]$ and the Jacobian, which means that the right-hand side of Eq. (15) needs to derive with respect to $\{J^\theta\}$. The critical current density is

$$J_c = \frac{J_{c0}}{\left(1+\sqrt{\alpha^2 B_\parallel^2 + B_\perp^2}/B_c\right)^\beta}, \tag{17}$$

where parameters $\alpha$, $\beta$, $B_c$ and the initial critical current density $J_{c0}$ is given in the calculation examples of section 3. Note that both the parallel and perpendicular components of the magnetic induction intensity $B_\parallel$ and $B_\perp$ are the function of $J^\theta$, and the calculation formula is shown in Appendix A. Then, an explicit-implicit hybrid algorithm (EIHA) is proposed, which means substituting $J_c$ at the moment $t$ into Eq. (15) and solving $\{J^\theta\}$ at the moment $t+\Delta t$ using an implicit algorithm. Since $J_c$ is constant at this time, the Jacobian $[J_{\text{jac}}]$ is

$$[J_{\text{jac}}] = -[C_1]^{-1}[L]^T[C_2][A_2]^{-1}[A_1]-[P], \tag{18}$$

where $[P]$ is a diagonal matrix and its main diagonal elements are [24]

$$P_{ii} = \frac{E_c n}{J_c}\left|\frac{J_i^\theta}{J_c}\right|^{n-1} \quad i=1,\ 2,\ \cdots,\ N. \tag{19}$$

The pseudocode of the explicit-implicit hybrid algorithm is shown in Algorithm 1. All the numerical simulations are performed on a PC [Intel(R) Core™ i9-12900K CPU @ 3.20 GHz, 32GB RAM, Windows 10 (64-bit)].

**Algorithm 1.** The explicit-implicit hybrid algorithm (EIHA) for the NI coils

---

Let $t = 0$. Set the time range $[0, t_{end}]$, the time step $\Delta t$, the iterative initial values of the hoop current density $\{J^\theta\}^{(0)} = \{J^\theta\}|_{t=0} = \{0\}$.

**while** $t < t_{end}$

  Calculate $J_c$ by Eq. (17);

  $t \leftarrow t + \Delta t$;

  Calculate $[J_{jac}]$ by Eq. (19);

  Calculate $\{J^\theta\}$ by Eq. (15) using the ode solvers in MATLAB;

**end**

$\{J^\theta\}$ for all times can be obtained, and $\{J^r\}$ can be calculated by the second equation of Eq. (10).

---

## 3 Results and discussions

*3.1 Validity and efficiency of the method*

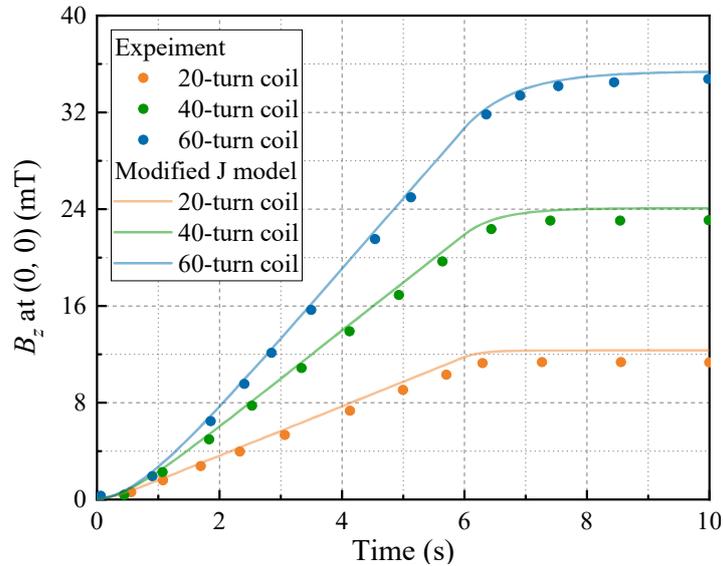

**Figure 1.** The comparison of magnetic field $B_z$ at point $(0, 0)$, and the results of the experiment are from [37].

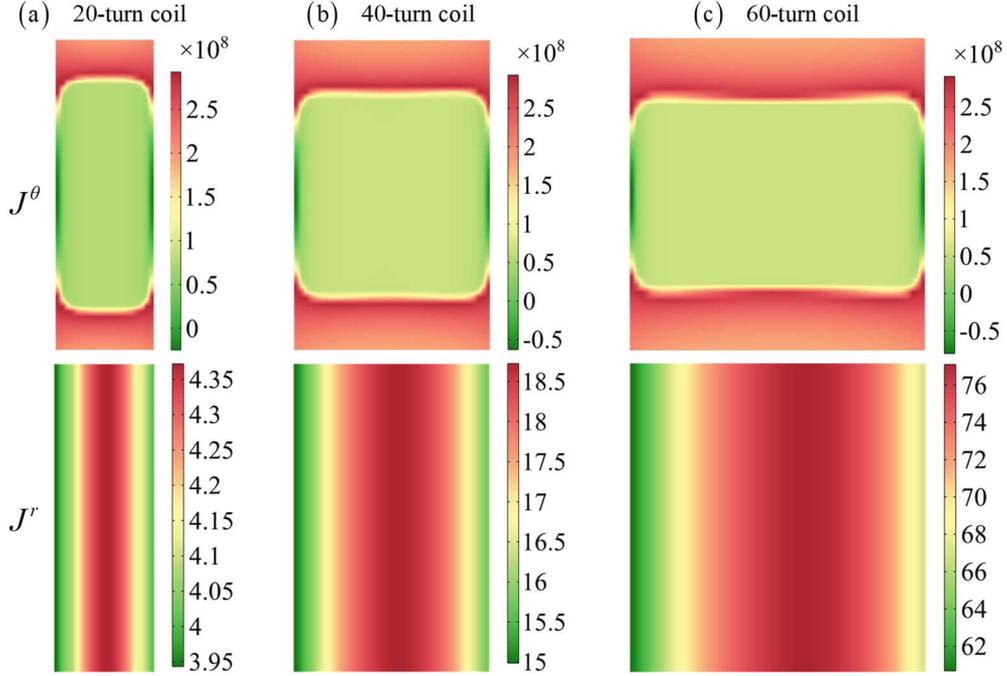

**Figure 2.** Hoop and radial current densities of (a) the 20-turn coil, (c) the 40-turn coil and (c) the 60-turn coil at the 10-th second.

The charging processes of three test coils are calculated, the numbers of turns are 20, 40 and 60 respectively and other parameters are shown in [37]. The target value of $I_{tr}$ and the charging rate are 30 A and 5 A/s, respectively. $N_w$ of the coils are respectively 60, 50 and 40, and the ode15s solver is used. In Fig. 1, it can be seen that the results of our method are in good agreement with the experimental data. Fig. 2 shows the hoop and radial current densities in these three coils at the 10-th second. The distribution of $\{J^\theta\}$ is very similar to that of the insulation coil, which means that it penetrates from the upper and lower ends of the coil to the middle. The radial current densities $\{J^r\}$ of the middle turns are larger than those of the inner and outer turns. In addition, the average hoop and radial currents are defined as [35]

$$I_{av}^\eta = \frac{1}{N_t}\sum_{i=1}^{N_t} I_i^\eta \qquad \eta = \theta \text{ or } r. \tag{20}$$

They are presented in Fig. 3 and compared with the results obtained by the equivalent circuit axisymmetric model [29] of the 20-turn and 60-turn coils, which are also in good agreement.

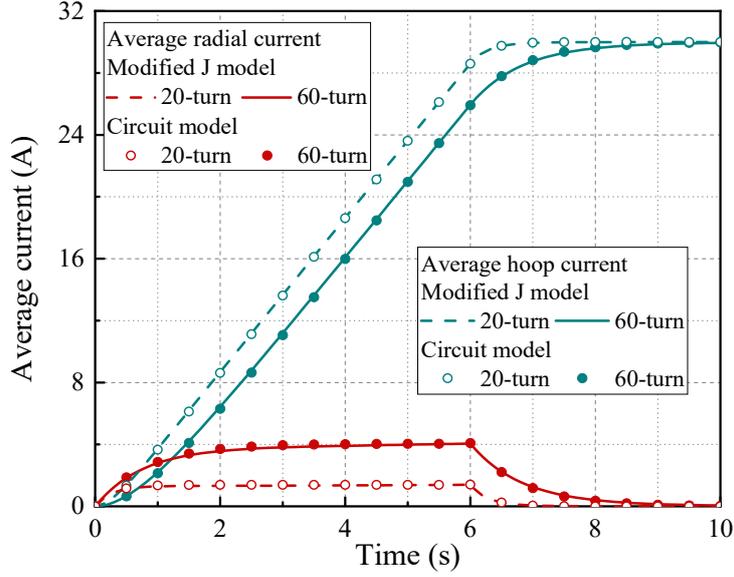

**Figure 3.** The average hoop and radial currents of the 20-turn and 60-turn coils.

The computation efficiency of stiffness solvers is compared, including ode15s (the backward differentiation formula (BDF) or the numerical differentiation formula (NDF) [38]), ode23s (the modified Rosenbrock formula [38]) and ode23t (the trapezoidal formula [39]), and EIHA can run fast and stably with large time steps of the order of $10^{-1}$ s. Table 1 presents the computation times using different solvers. It can be seen that ode15s has the fastest solution speed, while ode23t is almost the same. The efficiency of ode23s is the lowest but still acceptable. Meanwhile, the fully explicit algorithm is also tested by using ode23, ode45 and ode113. The explicit Runge-Kutta methods with different orders are used for ode23 and ode45 [40, 41], and the Adams-Bashforth-Moulton prediction-correction method is used for the ode 113 [38]. These solvers are all of variable step sizes in MATLAB, however we still need to specify the appropriate maximum allowed step sizes, otherwise the numerical results may be jittered or even NAN (i.e., not a number) due to the natural weakness of explicit algorithms for the stiff ODEs. For example, the computation times of ode23, ode45 and ode113 are respectively 48.9 s, 66.5 s and 47.3 s when $N_t = 20$ and $N_w = 25$, which are tens of times longer than EIHA's computation times. The above results do not contain the computation time of the coefficient matrixes $[M]$ and the magnetic field coefficient shown in Appendix A, as they are not time-expensive in these examples and only need to be calculated once.

**Table 1.** Computation times using different solvers (Unit: s)

|  | ode15s | ode23s | ode23t |
|---|---|---|---|
| 20 turns ($N_w = 25$) | 0.9 | 2.7 | 1.0 |
| 20 turns ($N_w = 60$) | 5.9 | 15.6 | 6.2 |
| 40 turns ($N_w = 25$) | 4.0 | 11.2 | 4.5 |
| 40 turns ($N_w = 50$) | 19.5 | 45.6 | 19.3 |
| 60 turns ($N_w = 25$) | 9.8 | 25.5 | 10.5 |
| 60 turns ($N_w = 40$) | 30.4 | 68.6 | 30.2 |

*3.2 Effect of magnetization and charging rates on loss*

**Table 2.** Parameters of the ReBCO tape at 77 K [42, 43].

| Parameter | Value |
|---|---|
| Tape width, $w$ | 4 mm |
| Tape thickness, $d$ | 0.1 mm |
| $J_{c0}$ @ 77 K and 0 T | $5.5 \times 10^8$ A/m$^2$ |
| $n$ value | 31 |
| $\alpha$, $\beta$ and $B_c$ in Eq. (16) | 0.0605, 0.758 and 103 mT |
| Contact resistivity, $\rho_{ct}$ | 70 $\mu\Omega \cdot$ cm$^2$ |

The hoop loss energy $Q_h$ and the radial loss energy $Q_r$ are respectively defined as the time integral of the hoop loss power $p_h$ and the radial loss power $p_r$ [35], i.e.,

$$\begin{cases} Q_h = \int_0^{t_{end}} p_h \, dt = \int_0^{t_{end}} dt \cdot \int_V E^\theta J^\theta dV \\ Q_r = \int_0^{t_{end}} p_r \, dt = \int_0^{t_{end}} dt \cdot \int_V E^r J^r dV \end{cases}, \quad (21)$$

where $V$ is the volume of the coils. A 60-turn coil is calculated to analyze the loss energies by different magnetization and charging rates. The inner radius is 30 mm and other parameters are shown in Table 2. During magnetization, the magnetic field is generated by a background magnet, and the target field at the point $(0, 0)$ is 1 T. As shown in Fig. 4, the hoop and radial currents are opposite. With the increase of the magnetization rate, their absolute values increase, leading to the higher loss energies of the hoop and radial directions. Due to the tiny superconducting resistance, the hoop loss energy is much less than that of the radial direction. The target current in the charging process is 60 A. As shown in Fig. 5, the higher current rate inducts the larger radial current, which leads to the larger radial loss energy. Because the hoop current is the main part during charging, the hoop loss energy is larger than that of the magnetization process.

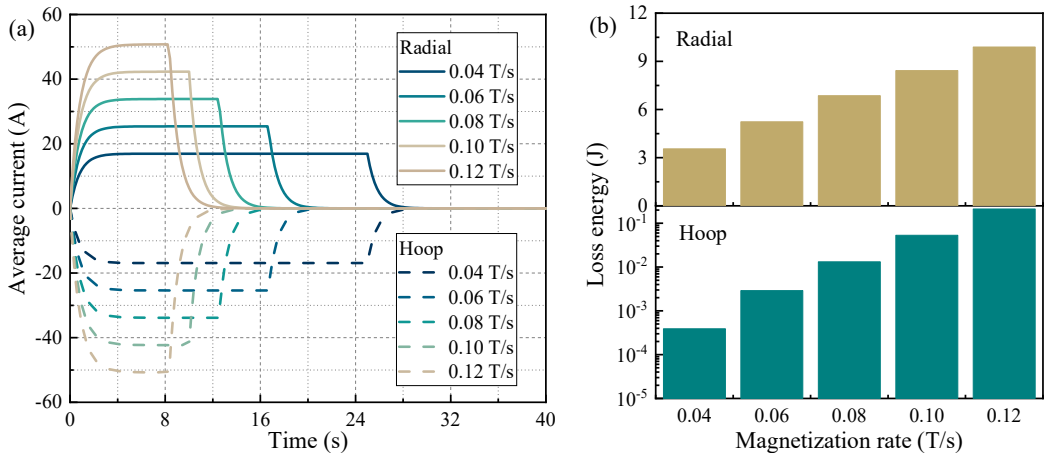

**Figure 4.** (a) The average hoop and radial currents and (b) loss energies by different magnetization rates.

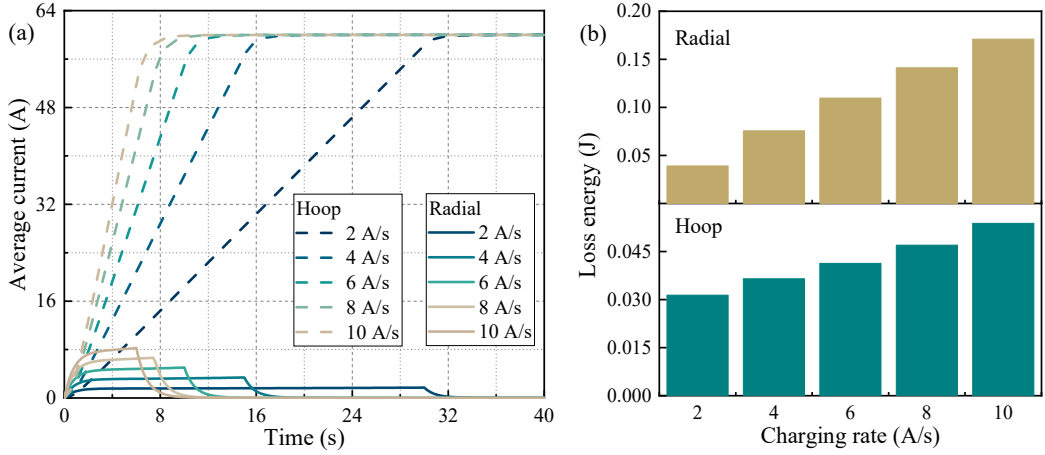

**Figure 5.** (a) The average hoop and radial currents and (b) loss energies by different charging rates.

*3.3 8×150-turn NI insert coil using homogenization model*

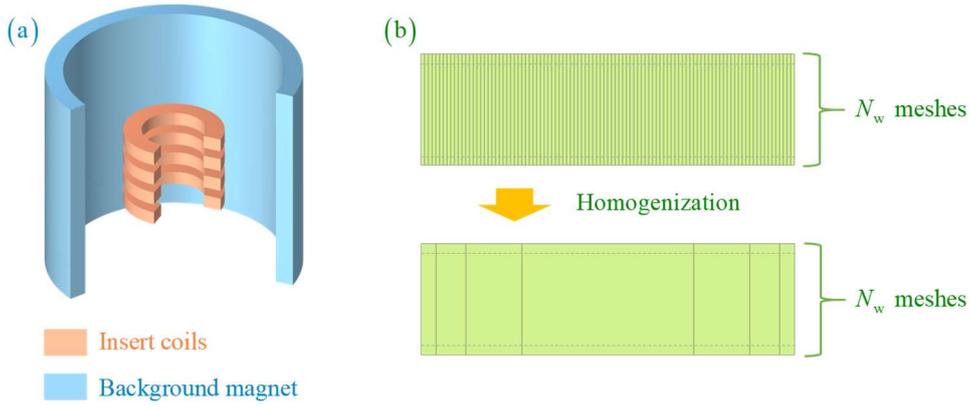

**Figure 6.** (a) A stack of NI pancake coils inserted in a background magnet, (b) the schematic diagram of the homogenization technique.

The common structure of the large magnet is a stack of NI pancake coils inserted in a background magnet, as shown in Fig. 6(a). Due to the large number of turns and the long magnetization and charging time, the numerical simulation of large magnets becomes a great challenge. The homogenization technique is an important measure to improve computational efficiency [44, 45], where the turns with similar electromagnetic fields are combined into a bulk, as shown in Fig. 6(b). For NI coils, due to merging turns, the integral path in Eq. (8) is the total perimeter and thickness of all turns in a bulk, i.e.,

$$(\nabla \Phi)^\theta \cdot \sum_i 2\pi R_{k,i} + N_{b,k} d\rho^r J^r = 0, \tag{22}$$

where $R_{k,i}$ is the radius of the *i*-th turn, and $N_{b,k}$ is the number of turns in the *k*-th bulk. Because $\{R_{k,i}\}$ is an arithmetic progression, Eq. (22) becomes

$$2\pi R_{b,k} \cdot (\nabla \Phi)^\theta + d\rho^r J^r = 0, \tag{23}$$

where $R_{b,k}$ is the radial coordinates of the midpoint of the *k*-th bulk. The current constraint equation becomes

$$N_{b,k} I_{tr} = \sum_i I_i^\theta + \sum_i I_i^r = \int_{S_{b,k}} J^\theta \, dS + \int_{D_{b,k}} J^r \, dD, \tag{24}$$

where $I_i^\theta$ and $I_i^r$ are respectively the hoop and radial currents of the *i*-th turn in the *k*-th bulk. The cross area $S_{b,k}$ and the total lateral area $D_{b,k}$ of the *k*-th bulk are respectively $N_{b,k} d w$ and $2\pi N_{b,k} R_{b,k} w$. The advantage of the homogenization technique is that it not only reduces the number of elements of $\{J^\theta\}$ and $\{J^r\}$, but also reduces the dimension of the coefficient matrixes, which are also time-consuming in the calculation of large magnets.

Table 3. Parameters of the 8×150-turn insert coil [42].

| Parameter | Value |
| --- | --- |
| Tape width, $w$ | 4 mm |
| Tape thickness, $d$ | 0.1 mm |
| $J_{c0}$ @ 4.2 K and 0 T | $1.8 \times 10^9$ A/m$^2$ |
| $n$ value | 31 |
| $\alpha$, $\beta$ and $B_c$ in Eq. (16) | 0.03813, 0.7122 and 631 mT |
| Contact resistivity, $\rho_{ct}$ | 70 $\mu\Omega \cdot$ cm$^2$ |
| Number of single pancake (SP) coils | 8 |
| Number of turns per coil | 150 |
| Inner radius | 20 mm |
| Gap between coils | 0.2 mm |

An 8×150-turn NI insert coil is calculated to show the computing efficiency of homogenization method, and parameters are shown in Table 3. The background magnetic field provides a non-uniform background field for 500 s, whose target value at the point $(0, 0)$ is 5 T. Charging process starts after 300 s with a charging rate of 0.25 A/s and a target value of 100 A. The whole process takes 1500 s. The number of meshes along the width is $N_w = 15$. Three homogenization ways are used, namely non-homogenization (NH), homogenization of 52 bulks (H1) and 34 bulks (H2). The numbers of turn in each bulk of H1 and H2 are respectively $\{1, 2, \underbrace{3 \cdots 3}_{48}, 2, 1\}$ and $\{1, 2, 3, 4, \underbrace{5 \cdots 5}_{26}, 4, 3, 2, 1\}$. Since the structure is symmetric along $z = 0$, only the 1/2 region needs to be calculated. The element of $[M]$ in this case can be obtained by the mirror image method as follows [44]:

$$M_{\text{sym}, ij}(r_i, z_i, r_j, z_j) = M_{ij}(r_i, z_i, r_j, z_j) + M_{ij}(r_i, z_i, r_j, -z_j). \quad (25)$$

The loss power is shown in Fig. 7, and it can be seen that the loss is almost entirely radial in the magnetization process, which is similar to the discussion in Section 3.3. During charging process, the contribution of the hoop loss power to the total is obvious, which is due to the slow charging rate. Fig. 8 shows the hoop and radial current densities of the 8×150 turn coil and the results of different homogenization ways are almost the same. Since the innermost and outermost turns of each SP are not merged, the end effects of the current distribution are well preserved. The total loss energy is defined as $Q = Q_h + Q_r$ and is shown in Table 4 by different homogenization ways. It can be seen that the relative error increases as the number of bulks decreases slightly. Importantly, the times to compute both the coefficient matrix and EIHA are greatly reduced, which makes it possible to simulate the electromagnetic fields of the high-field large-scale NI magnets.

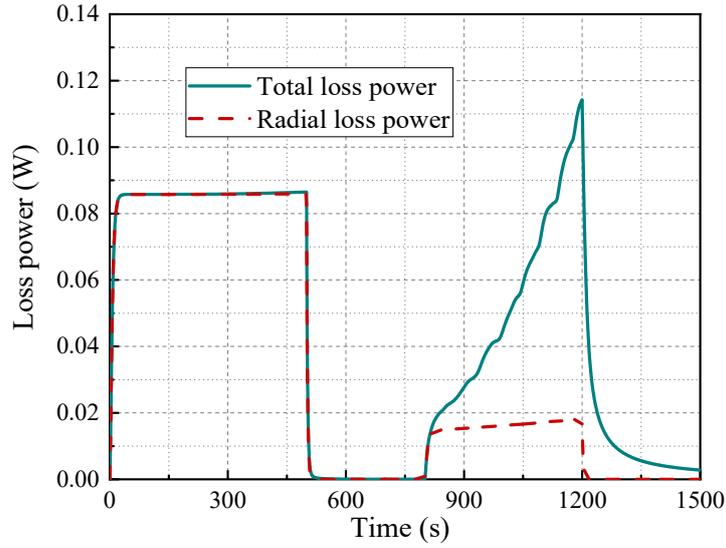

**Figure 7.** (a) The hoop and (b) radial loss powers of the 8×150-turn insert coil.

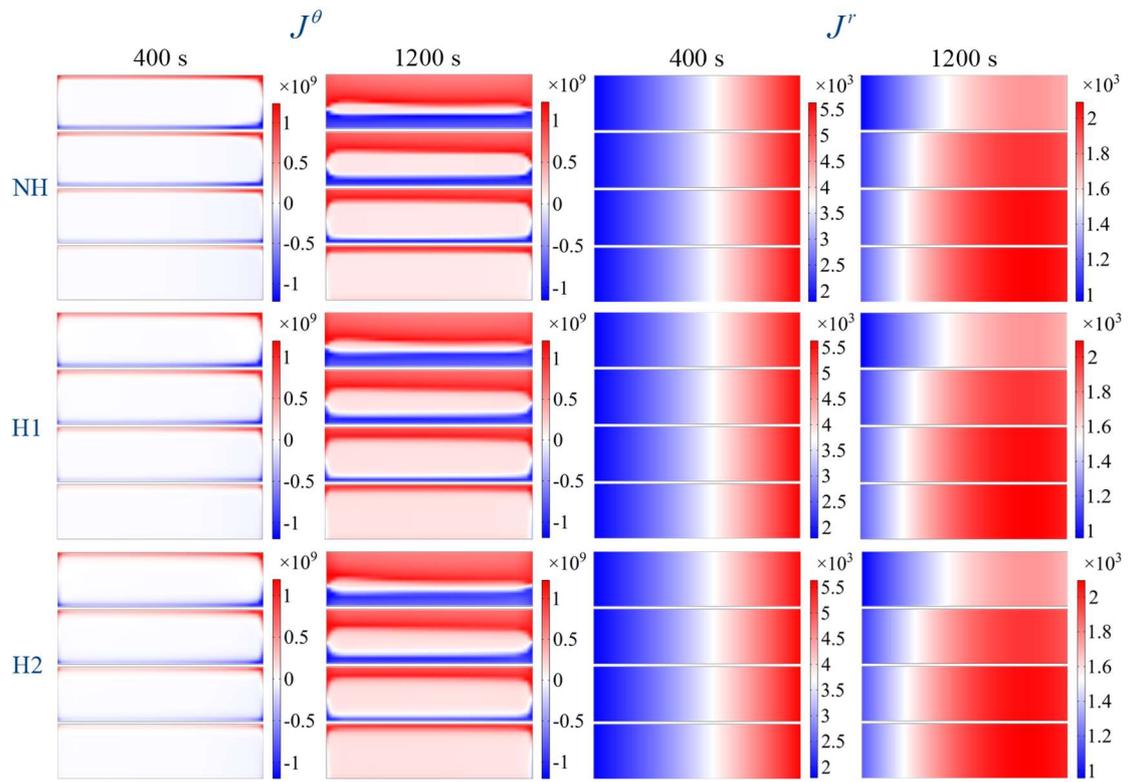

**Figure 8.** Hoop and radial current densities of the 8×150-turn NI insert coil with different homogenization ways.

**Table 4.** The total loss energy and computing time by different homogenization ways. (CM means the coefficient matrixes.)

|  | NH | H1 | H2 |
| --- | --- | --- | --- |
| Total loss energy, $Q$ | 67.02 J | 66.7 J | 66.43 J |
| Relative error | - | 0.48% | 0.88% |
| Computing time of CM | 1.95 h | 0.04 h | 0.01 h |
| Computing time of EIHA | 15.74 h | 1.42 h | 0.6 h |
| Total computing time | 17.69 h | 1.46 h | 0.61 h |

*3.4 Insulation coils*

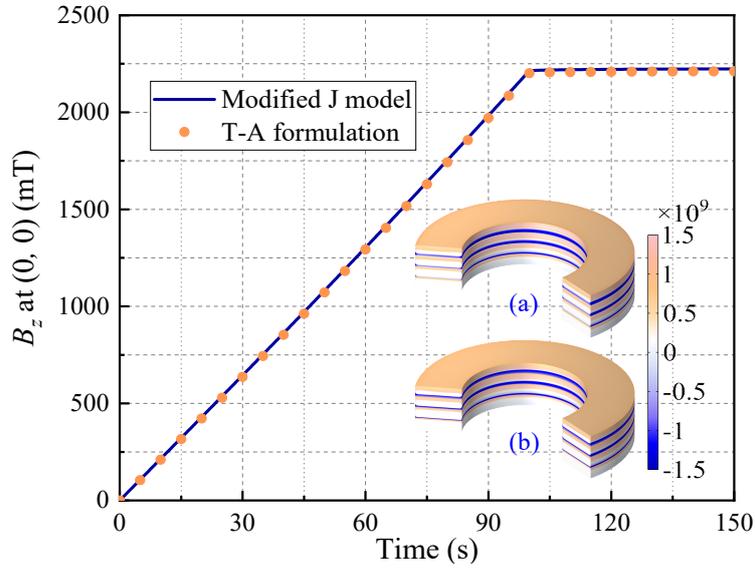

**Figure. 9** The magnetic field $B_z$ at the point $(0, 0)$ of an insulation 8×150-turn coil using the modified J model with a larger contact resistivity and T-A formulation. In the illustration, (a) and (b) are respectively the hoop current density at 100 s obtained by modified J model and T-A formulation (Only the upper half of the coil is shown).

The insulation coil can also be calculated by the modified J model by setting a large contact resistivity. The charging process of an insulation 8×150-turn coil is used to test our method, which is compared by T-A formulation. Here, the contact resistivity is $10^9$ $\mu\Omega \cdot cm^2$, and other parameters are shown in Table 3. The transport current is $I_{tr} = t$ and the target value is 100 A. The magnetic field at the point $(0, 0)$ estimated by our method is shown in Fig. 9 and the field delay phenomenon is not seen, which is the characteristic of an insulation coil. The results of the field and the current density are in good agreement with T-A formulation. Note that the modified J model has better convergence even for a large contact resistivity.

## 4 Conclusion

A modified J model is proposed in this paper to analyze the electromagnetic field of the NI coils. In order to avoid the inefficiency of the explicit algorithm caused by the stiffness of the equations, the explicit-implicit hybrid algorithm is designed. The modified J model is not time-expensive and its results are in good agreement with the experiments and the equivalent circuit model. Then, the loss powers of NI coils are studied with the modified model, and the homogenization model is developed for large-scale NI magnets. It is proved by an example that the homogenization model has a good accuracy and a relatively fast computation speed. The modified J model can be used to simulate the insulation coils by adjusting the contact resistivity.

At present, our method is only for NI magnets with axisymmetric structures. In three-dimensional (3D) structures, the current continuity equation is required to be satisfied at any point. Our future work is to extend the modified model for 3D cases.

## Acknowledgments

The authors acknowledge the supports from the National Natural Science Foundation of China (Nos. U2241267, 11872195, 11932008 and 12302278), the Fundamental Research Funds for the Central Universities (No. lzujbky-2022-48).

**Appendix A The magnetic field produced by the hoop current density [17]**

The parallel and perpendicular magnetic fields produced by the hoop current density are

$$\begin{cases} B_{\parallel,i} = B_i^r = \dfrac{\mu}{2\pi} \sum_{j=1}^{N} \dfrac{(z_i - z_j) \cdot \Delta S_j J_j^\theta}{r_i \sqrt{(z_i - z_j)^2 + (r_i + r_j)^2}} \left[ -K + \dfrac{(z_i - z_j)^2 + r_i^2 + r_j^2}{(z_i - z_j)^2 + (r_i - r_j)^2} \cdot F \right] \\ B_{\perp,i} = B_i^z = \dfrac{\mu}{2\pi} \sum_{j=1}^{N} \dfrac{\Delta S_j J_j^\theta}{\sqrt{(z_i - z_j)^2 + (r_i + r_j)^2}} \left[ K + \dfrac{r_j^2 - (z_i - z_j)^2 - r_i^2}{(z_i - z_j)^2 + (r_i - r_j)^2} \cdot F \right] \end{cases}, \quad (A1)$$

where *K* and *F* are respectively the first and second complete elliptic integrals.